# Test Case Generation using Mutation Operators and Fault Classification

Mrs. R. Jeevarathinam [#1], Dr. Antony Selvadoss Thanamani [*2]

[#] *Department of Computer Science*

*SNR Sons College, Coimbatore, Tamilnadu, India.*

[1]

[*] *Associate Professor and Head*

*Department of Computer Science*

*NGM College, Pollachi, Tamilnadu, India.*

[2]

*Abstract*— Software testing is the important phase of software development process. But, this phase can be easily missed by software developers because of their limited time to complete the project. Since, software developers finish their software nearer to the delivery time; they don't get enough time to test their program by creating effective test cases. . One of the major difficulties in software testing is the generation of test cases that satisfy the given adequacy criterion Moreover, creating manual test cases is a tedious work for software developers in the final rush hours. A new approach which generates test cases can help the software developers to create test cases from software specifications in early stage of software development (before coding) and as well as from program execution traces from after software development (after coding). Heuristic techniques can be applied for creating quality test cases. Mutation testing is a technique for testing software units that has great potential for improving the quality of testing, and to assure the high reliability of software. In this paper, a mutation testing based test cases generation technique has been proposed to generate test cases from program execution trace, so that the test cases can be generated after coding. The paper details about the mutation testing implementation to generate test cases. The proposed algorithm has been demonstrated for an example.

## I. INTRODUCTION

Software testing is a standard method of assuring software quality. Software testing is an important activity to assure the quality of software. Unfortunately, software testing is very labor intensive and very expensive. It can take about 50 percents of total cost in software developing process [1]. The software testers may need to spend a longer time using many test cases if the test data used are not of high quality. Therefore, a performance of executing test case is an important issue to reduce the testing time. Software testing is usually the first part of software development stages, which software developers decide to omit when there is a limited time to deliver the software. In other word, developers may not have enough time after they finished their coding to create test cases to test their code. Generating test cases can resolve these problems. This not only helps developers to test their program when they finish coding but also controls the developers to program the software as defined in the software specification [2]. In this case the software specifications are the main sources for generating test cases as these documents describe the software system to be developed in detail.

One of the most difficult and expensive parts of applying these techniques has been the actual generation of test data- which has traditionally been done by hand

The general aim of the research reflected in this paper is to formalize, and mechanize where possible, routine aspects of testing. Such formalization has two benefits. First, it makes it easier to analyse a given test set to ensure that it satisfies a specified coverage criterion. Second, it frees the test engineer to concentrate on less formalization, and often more interesting tests. Developers have responded to this need in many ways, including improving the process, increasing the attention on early development activities, and using formal methods to describe requirements, specifications, and designs. Although all of these improvements help create software that is of higher quality and higher reliability, the software still needs to be tested, and the more stringent needs for the product also means that the testing method must be more effective at finding problems in the software. Project and test managers are more than ever in a position where they need solid information for how to apply scarce resources. Applying structured, precisely defined testing techniques allows development resources to be used more wisely.

Specification-based testing refers to creating test inputs from the software specifications. Specification-based testing allows tests to be created earlier in the development process, and be ready for execution before the program is finished. Additionally, when the tests are generated, the test engineer will often find inconsistencies and ambiguities in the specifications, which allows problems to be found and eliminated early. Specifications can be used as a basis for output checking, which significantly reduces one of the major costs of testing. Another advantage is that the essential part of the test data can be independent of any particular implementation of the specifications. Specification-based testing is also important for conformance testing, where access to the code is not provided, but specifications for the product are.

There is an increasing need for effective testing of software for growing applications, such as web applications and e-commerce require software that exhibits more reliability than most traditional application areas. Without software that functions reliably, businesses that operate on the web will lose money, sales, and customers. In recent years, the phrase "fault-based testing" has been applied to techniques that choose test data that attempt to show the presence (or absence) of specific faults during unit testing. Techniques have been developed that determine whether test data can detect specific faults (i.e., mutation analysis [3]), and the theoretical properties of fault-based testing have been explored [5, 4].

The mutation testing is a fault based testing strategy that measures the quality of testing by examining whether the test set, test input data used in testing can reveal certain type of faults. Mutation testing helps testers create test data by interacting with them to strengthen the quality of the test data. Faults are introduced into programs by creating many versions of the software, each containing one fault. Test cases are used to execute these faulty programs with the goal of causing each faulty program to produce incorrect output (fail). Hence the term mutation; faulty programs are mutants of the original, and a mutant is killed when it fails. When this happens, the mutant is considered dead and no longer needs to remain in the testing process because the faults represented by that mutant have been detected.

## II. RELATED WORK

If testers want to test functional requirements, they may use black-box testing technique. Black-box testing [6] does not need knowledge of how software is programmed. Test oracles are specified by software design or software specifications. Testers inject test data to execute program, then compare actual result with the specified test oracle. By contrast, white-box testing needs knowledge of how software is programmed. In white-box testing, paths or statements which has been executed are test oracle. These are called coverage criteria. There are three main types of coverage criteria: statement, coverage, branch coverage, and path coverage. Statement coverage reports whether each statement is encountered by the test suite or not. Branch coverage reports whether every branch structure (if – else clause or while clause) has been executed for true and false condition in each branch. Finally, path coverage reports whether all possible paths in function has been tested.

In Object-oriented context, the structure of software is more complicated than the structural one. Conventional test approaches may not be enough for testing. The combination of those two traditional approaches is called Gray-box testing [7]. In Gray-box testing, test data generates based on the high level design which specifies the expected structure and behaviour of system. Gray-box testing investigates the coverage criteria of white-box method and finds all possible coverage paths. In addition, the generated test case should be satisfied with functional requirement as in the black-box testing criteria.

Many automated test case generation techniques produce test cases based on Gray-box method. Not only does Gray-box testing concern functional requirement as black box testing, but also concerns on behaviours of system. Clarke [8] proposed an empirical study which compared efforts between automate test generation and manual test generation. In his report test data was generated from extended finite state model (EFSM). The research shows that the automate test data generation could reduce an effort from manual test data generation for more than 88 percents. Xu and Yang [9] proposed test data generation framework called JMLAutoTest framework. JMLAutoTest framework generates test data from Java Modelling Language (JML) [[10] [11]]. JML is a notation for specifying behaviour and interface in Java class and method. Since JML is a formal specification, developers should spend efforts to understand JML before writing specification. Because UML diagrams are now widely used for software development [12], generating test data from UML diagrams should help developer to reduce a great number of efforts.

Wang, et.al [13] proposed test data generation from activity diagram. They extracted a test scenario from activity diagram. The test scenario is a sequence of possible paths in activity diagram. From these paths, the executing sequence of program has been generated in order to cover all possible paths. However, activity diagram describes flows of system, not the behaviour of the system. Due to performance of generating test data and a concern of size of test data set, heuristic techniques are applied for test data generation. GADGET [14] and TGEN [15] use genetic algorithm to improve quality of generating test data. GADGET generates test data from a control flow graph generated from source code. A fitness function is defined for each condition node in control flow graph. An empirical study showed that test data generated by GADGET covers more than 93 percents of source code, while random testing achieves around 55 percents. TGEN transforms a control flow graph to a control dependency graph (CDG). Each part of CDG represents the smallest set of predicate to traverse every node in control flow graph. Both GADGET and TGEN generate test data using white box method; therefore, test data can be generated only after software is finished. Using Genetic algorithm to generate test data from software model is proposed in [16]. JML is a model for generating test data. Fitness function is calculated by coverage of paths and post condition defined by JML.

Because of the large number of mutant programs that must be generated and run, early designers of mutation analysis systems considered individually creating, compiling, linking, and running each mutant more difficult, and slower, than using an interpretive system [[17] [18]] . It was considered likely that the cost of compiling large numbers of mutants would be prohibitive. Of the interpreter-based systems that have been developed, Mothra is the most recent and comprehensive [[19] [21]]. In these conventional, interpreter-based mutation analysis systems, the source code is translated into an internal form suitable for interpretive execution and mutation. For each mutant, a mutant generator program produces a "patch" that, when applied to the internal form, creates the desired alternate program. The translated program plus the collection of patches represents a program neighbourhood. To run a mutant against a test case, the interpreter dynamically applies the appropriate patch and interpretively executes the resulting alternate internal form program. A number of attempts to overcome the performance problem have been made. Some approaches attempt to limit the number of mutants that must be run. In selective mutation [26], only a subset of the possible mutagens is used, resulting in fewer mutants being created. Preliminary results suggest that selective mutation may provide almost the same test coverage as non-selective mutation under certain conditions. Running only a sample of the mutants [27] has also been suggested.

In extreme cases, however, it is necessary to run almost all the mutants. In other approaches, the use of non-standard computer architectures has been explored. Unfortunately, full utilization of these high performance computers requires an awareness of their special requirements as well as adaptation of software. Work has been done to adapt mutation systems to vector processors [23], to SIMD [22] and hypercube (MIMD) machines [[24] [25]]. However, it is the very fact that these architectures are non-standard that limits the appeal of these approaches. Not only are they not available in most development environments, but testing software designed for one operational environment (machine, operating system, compiler, etc.) on another is fraught with risks.

The approaches above do not squarely address the primary factor that causes conventional systems to be slow: interpretative execution. As noted previously, the overhead of compiling many mutant programs outweighs the benefit of increased execution speed. Compiler-integrated [20] program mutation seeks to avoid excessive compilation overhead and yet retain the benefit of compiled speed execution. In this method, the program under test is compiled by a special compiler. As the compilation process proceeds, the effects of mutations are noted and code patches that represent these mutations are prepared. Execution of a particular mutant requires only that the appropriate code patch be applied prior to execution. Patching is inexpensive and the mutant executes at compiled-speeds. Unfortunately, crafting the needed special compiler is an expensive undertaking. Modifying an existing compiler reduces this burden somewhat, but the task is still technically demanding. Moreover, for each new computer and operating system environment, this task must be repeated.

## III. FAULT CLASSIFICATION

A test case that distinguishes the program from its mutant is considered to be effective at finding faults in the program. The effectiveness of mutation testing, like other fault-based approaches, depends heavily on the types of faults that the mutation system is designed to represent. Since mutation testing uses mutation operators to implement faults, the quality of the mutation operators is crucial to the effectiveness of mutation testing. Although mutation testing has a rich history, most mutation operators have been developed for procedural programs. OO languages contain new features such as encapsulation, inheritance, and polymorphism. These features introduce the potential for new faults. Therefore, existing mutation operators for procedural programming languages are not sufficient for programs written in OO languages and new OO-specific language operators are needed.

The effectiveness of mutation testing depends heavily on the types of faults that may be represented. In these new kinds of faults, some of which are not modelled by traditional mutation operators. Which are insufficient to test these OO language features, particularly at the class testing level. This paper introduces a new set of class mutation operators for the OO languages. These operators are based on specific OO faults and can be used to detect faults involving inheritance, polymorphism, and dynamic binding, thus are useful for inter-class testing. The faults modelled by these operators are not general; they can be application-specific or programmer-specific. Therefore, to execute mutation testing with these operators, they should be selected based on the characteristic of the program to be tested. The previous attempts suffered from not having a general fault model. The previous OO mutation operators do not handle several fault types and did not handle all OO features. Faults can be classified as occurring at the intra-method level, inter-method level, intra-class level, and inter-class level.

Intra-method level faults occur when the functionality of a method is implemented incorrectly. A method in a class corresponds to the unit of the conventional program testing. Inter-method and intra-class level faults are made at the interactions between pairs of methods of a single class or between pairs of methods that are not part of a class construct in non-OO languages. Because methods are getting smaller and interactions among methods are increasingly encoding the design complexity. Inter-class level faults include faults that occur due to the object-oriented specific features such as encapsulation, inheritance, polymorphism, and dynamic binding.

## IV. MUTATION OPERATORS

There are three kinds of mutation operators available namely statement level operators, method level operators and class level operators.

Statement level mutation operators involve the creation of a set of mutant programs of the program being tested. Each mutant differs from the original program by one mutation. A mutation is a single syntactic change that is made to a program statement.
Operand Replacement Operators (ORO) - Replacing a single operand with another operand or constant.
Expression Modification Operators (EMO) – Replacing an operator or inserting a new operator.
Statement Modification Operators (SMO) – Replacing or deleting a statement or part of the statement.

Method level mutation operators are used in unit and integration level testing and can be classified into two levels: (1) intra-method, (2) inter-method. This classification follows definitions by Harrold and Rothermel [25] [27] and Gallagher and Offutt [26] [23].

Intra-method level faults occur when the functionality of a method is implemented incorrectly. Testing within classes corresponds to unit testing in conventional programs. So far, researchers have assumed that traditional mutation operators for procedural programs will suffice for this level (with minor modifications to adapt to new languages).

Inter-method level faults are made on the connections between pairs of methods of a single class. Testing at this level is equivalent to integration testing of procedures in procedural language programs. Interface mutation which evaluates how well the interactions between various units have been tested, is applicable to this level.

Class level mutation operators can be classified into two levels: (1) intra-method, (2) inter-method.

Intra-class testing is when tests are constructed for a single class, with the purpose of testing the class as a whole. Intra-class testing is a specialization of the traditional unit and module testing. It tests the interactions of public methods of the class when they are called in various sequences. Tests are usually sequences of calls to methods within the class, and include thorough tests of public interfaces to the class.

Inter-class testing is when more than one class is tested in combination to look for faults in how they are integrated. Inter-class testing specializes the traditional integration testing and seldom used subsystem testing, where most faults related to polymorphism, inheritance, and access are found.

Based on the fault classification, Ma et al. [28] developed a comprehensive set of class mutation operators for Java. There are 24 mutation operators explained below. Each mutation operator is related to one of the following six language feature groups. The first four groups are based on language features that are common to all object oriented languages. The fifth group includes language features that are Java-specific, and the last group of mutation operators are based on common object oriented programming mistakes. As is usual with mutation operators, they are only applied in situations where the mutated program will still compile.

### A. Information Hiding

Access control is one of the common sources of mistakes among object oriented programmers. The semantics of the various access levels are often poorly understood, and access for variables and methods is not always considered during design. Poor access definitions do not always cause faults initially, but can lead to faulty behaviour when the class is integrated with other classes, modified, or inherited from. The Access Control mutation operator, Access modifier change (AMC) has been developed for this category.

### B. Inheritance

Although inheritance is a powerful and useful abstraction mechanism, incorrect use can lead to a number of faults. Seven mutation operators have been defined to test the various aspects of using inheritance, covering variable hiding, method overriding, the use of super, and definition of constructors and are listed below.

IHD-Hiding variable deletion     IHI-Hiding variable insertion
IOD-Overriding method deletion   IOP- Overriding method calling position change
IOR-Overriding methods rename    ISK-Super keyword deletion
IPC-Explicit call of a parent's constructor deletion

### C. Polymorphism

Polymorphism and dynamic binding allow object references to take on different types in different executions and at different times in the same execution. That is, object references may refer to objects whose actual types differ from their declared types. In most languages (including Java and C++), the actual type can be any type that is a subclass of the declared type. Polymorphism allows the behaviour of an object reference to differ depending on the actual type. Four operators have been developed for this category.

PNC- new method call with child class type
PMD- Instance variable declaration with parent class type
PPD -Parameter variable declaration with child class type
PRV- Reference assignment with other comparable type

### D. Overloading

Method overloading allows two or more methods of the same class or type family to have the same name as long as they have different argument signatures. Just as with method overriding (polymorphism), it is important for testers to ensure that a method invocation invokes the correct method with appropriate parameters. Four mutation operators have been defined to test various aspects of method overloading.

OMR- Overloading method contents change
OMD- Overloading method deletion
OAO- Argument order change
OAN- Argument number change

### E. Java Specific Features

Because mutation testing is language dependent, mutation operators need to reflect language-specific features. Java has a few object-oriented language features that do not occur in all object oriented languages and four operators have been defined to ensure correct use of these features. They cover use of this, static, default constructors and initialization.

JTD- this keyword deletion
JSC- static modifier change
JID- Member variable initialization deletion
JDC-Java-supported default constructor creation

### F. Common Programming Mistakes

This category attempts to capture typical mistakes that programmers make when writing object oriented software. These are related to use of references and using methods to access instance variables. Four operators have been developed for this category.

EOA- Reference assignment and content assignment replacement

EOC- Reference comparison and content comparison replacement
EAM- Accessor method change
EMM- Modifier method change

## V. RELATIONSHIP BETWEEN FAULTS AND OPERATORS

The following table relates the fault types and our mutation operators. All faults are covered, and some required multiple mutation operators. Conversely, some of the mutation operators cover more than one fault

| Faults | Class Mutation Operators |
|---|---|
| State visibility anomaly | IOP |
| State definition inconsistency (due to state variable hiding) | IHD, IHI |
| State definition anomaly (due to overriding) | IOD |
| Indirect inconsistent state definition | IOD |
| Anomalous construction behaviour | IOR, IPC, PNC |
| Incomplete construction | JID, JDC |
| Inconsistent type use | PID, PNC, PPD, PRV |
| Overloading methods misuse | OMD, OAO, OAN |
| Access modifier misuse | AMC |
| Static modifier misuse | JSC |
| Incorrect overloading-methods implementation | OMR |
| Super keyword misuse | ISK |
| This keyword misuse | JTD |
| Faults from common programming mistakes | EOA, EOC, EAM, EMM |

## VI. EFFECTIVENESS OF MUTATION OPERATORS

TCAS/Siemens has an internal state which is large relative to the number of inputs and outputs. TCAS, aircraft collision avoidance, is a part of a set of C programs that came originally from Siemens Corporate Research and was subsequently modified by Rothermel and Harrold [26]. These programs are used in research on program testing, so they come with extensive test suites and sets of faulty versions. There are 12 input variables specifying parameters of own aircraft and another aircraft and one output variable, alt_sep, a resolution advisory to maintain safe altitude separation between the two aircrafts. The program computes intermediate values and prints alt_sep to the standard output. The program has minimal documentation, and we wrote a formal specification for it. The following table gives the results of various mutation operators in terms of number of mutants generated, number of traces produced and the percentage of fault coverage.

There are several issues that need to be considered to evaluate the usefulness and effectiveness of the OO class level mutation operators. First is the issue of equivalent mutants. Equivalent mutants do not affect the semantics of the program; therefore they are useless for mutation testing. Second, some operators can generate mutants that are easily killed. Although mutants make simple syntactic changes to the program, their semantic impacts can vary greatly. The impacts of OO operators on the semantics can vary from affecting the method to the semantics of the entire class. For example, the IOD operator swaps overriding methods with its parent's. The effect of IHD, on the other hand, extends over the whole class because it handles instance variable, which determine the class state. It is possible that some of these mutation operators will create mutants that are too easily killed. Finally, the mutation operators need to be evaluated in terms of their effectiveness of detecting faults in OO programs. The AMC and JSC operators produced a lot of equivalent mutants, and the PNC, PMD, PPD and IHI operators produced equivalent mutants when overriding was present.

TABLE I
RESULT OF MUTATION OPERATORS AND THEIIR FAULT COVERAGE

| Operator | No.of Mutants | No.of Traces | Coverage |
|---|---|---|---|
| AMC | 202 | 24 | 96.6% |
| IOD | 72 | 21 | 87.9% |
| ISK | 130 | 21 | 93.1% |
| IHD | 116 | 14 | 62.9% |
| PNC | 74 | 18 | 94.2% |
| PPD | 72 | 21 | 90.7% |
| PMD | 144 | 29 | 83.7% |
| JTD | 12 | 4 | 52.4% |
| JSC | 83 | 17 | 85.2% |
| IHI | 97 | 22 | 76.4% |

## VII. CONCLUSION

This paper presents a comprehensive set of mutation operators to test for faults in the use of object-oriented features. These mutation operators are based on an exhaustive list of OO faults, which gives them a firm theoretical basis. As a result, they correct several problems. These mutation operators are designed with an emphasis on the integration aspects of Java to support interclass level testing, and will help testers find faults with the use of language features such as access control, inheritance, polymorphism and overloading. Thus, this provides a way to improve the reliability of OO software.

AUTHORS PROFILE

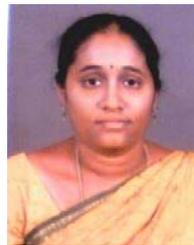

**Mrs R. Jeevarathinam**  graduated with MCA in 2001 from Bharathiar University, India and completed M.Phil from Bharathidasan University, India during 2003-04. Her areas of Interest include Software Engineering  & Data Mining. She has about 8 years of teaching experience. Currently she is working as a Sr. Lecturer in CS department at SNR Sons College, Coimbatore, India and also pursuing PhD of Mother Teresa Women University, India. She has published a number of papers in various national & international journals & conferences. She is an active IEEE student member

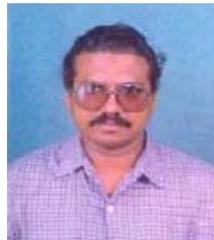

**Dr.Antony Selvadoss Thanamani** is presently working as Reader in the Dept of Computer Science, NGM College, India. He has published more than twenty papers in national/journals and more than ten books. His areas of interest includes E-Learning, Software Engineering, Data Mining, Networking and etc. He has about 20 years of teaching experience. He is guiding many research scholars and has published many papers in national and international conference and in many international journals